\documentclass[twocolumn,superscriptaddress]{revtex4-1}
\usepackage{graphicx}
\usepackage{dcolumn}
\usepackage{bm}

\begin{document}

\title{Power and spectral characterization of photonic integrated circuit based axicon like lens}

\author{Rijan Maharjan}
\email{rm@pinstitute.org}
\affiliation{Biophotonics Lab, Phutung Research Institute, PO Box 12335, Kathmandu 44600, Nepal}
\author{Sanket Bohora}
\affiliation{Biophotonics Lab, Phutung Research Institute, PO Box 12335, Kathmandu 44600, Nepal}
\author{Pravin Bhattarai}
\affiliation{Biophotonics Lab, Phutung Research Institute, PO Box 12335, Kathmandu 44600, Nepal}
\author{Iain Crowe}
\affiliation{Photon Science Institute, Department of Electrical and Electronic Engineering, University of Manchester, Manchester M13 9PL, UK}
\author{Richard J. Curry}
\affiliation{Photon Science Institute, Department of Electrical and Electronic Engineering, University of Manchester, Manchester M13 9PL, UK}
\author{Richard Hogg}
\affiliation{Electronics and Nanoscale Engineering, University of Glasgow, Glasgow G12 8QQ, UK}
\author{David Childs}
\affiliation{Electronics and Nanoscale Engineering, University of Glasgow, Glasgow G12 8QQ, UK}
\author{Ashim Dhakal}
\affiliation{Biophotonics Lab, Phutung Research Institute, PO Box 12335, Kathmandu 44600, Nepal}

\begin{abstract}
We demonstrate an on-chip Silicon-on-Insulator (SOI) axicon etched using a low resolution (200 nm feature size, 250 nm gap) deep-ultraviolet lithographic fabrication. The axicon consists of circular gratings with seven stages of 1x2 multimode interferometers. We present a technique to apodize the gratings azimuthally by breaking up the circles into arcs which successfully increased the penetration depth in the gratings from $\approx$5 $\mu$m to $\approx$55 $\mu$m. We characterize the device's performance by coupling 1300$\pm$50 nm swept source laser in to the chip from the axicon, and measuring the out-coupled light from a grating coupler. Further, we also present the implementation of balanced homodyne detection method for the spectral characterization of the device and show that the position of the output lobe of the axicon does not change significantly with wavelength.

\end{abstract}

\maketitle
%%%%%%%%%%%%%%%%%%%%%%%%%%  body  %%%%%%%%%%%%%%%%%%%%%%%%%%
\section{Introduction}
\label{sec:introduction}
Increasing utility of smaller and cheaper electro-optical devices has necessitated the development of various on-chip photonic devices in recent years. Integration of multiple devices into a single photonic integrated circuit (PIC) could not only make such devices smaller in size, but also increase accessibility owing to the lower cost of fabrication. One such potentially useful device is an axicon. A traditional axicon is a lens with a conical face on one side, and a planar face on other. An axicon can convert any gaussian beam into a profile similar to a bessel beam, and thus can be useful for technologies like optical traps \cite{GarcesChavez2002}. Axicons have also been used in imaging techniques like optical coherence tomography to improve the depth of focus compared to more traditional focusing solutions \cite{Leitgeb2006, Ding2002}. However, traditional axicons are not suitable for miniaturization in chip-based devices and as such, a chip-based axicon could prove to be beneficial for PICs.

In this letter, we demonstrate an on-chip SOI axicon etched using a low resolution deep-ultraviolet (DUV) lithographic fabrication. The axicon consists of circular gratings with seven stages of 1x2 multimode interferometers. Devices with circular structure have been demonstrated in context of focusing grating couplers, although spectral characterization of such structures has so far been incomplete~\cite{Doerr2011}. Here we present a methodology to characterize the spectral performance of such devices as well. Moreover, we present a technique to apodize the gratings azimuthally that is compatible with a low resolution fabrication technique, which allows deeper light penetration depth in the device - a necessity for an axicon.

\section{Materials and methods}
\label{sec:methods}

% actual device description - size, periods, apodization
\begin{figure}[htbp]
\centering
\includegraphics[width=0.9\linewidth]{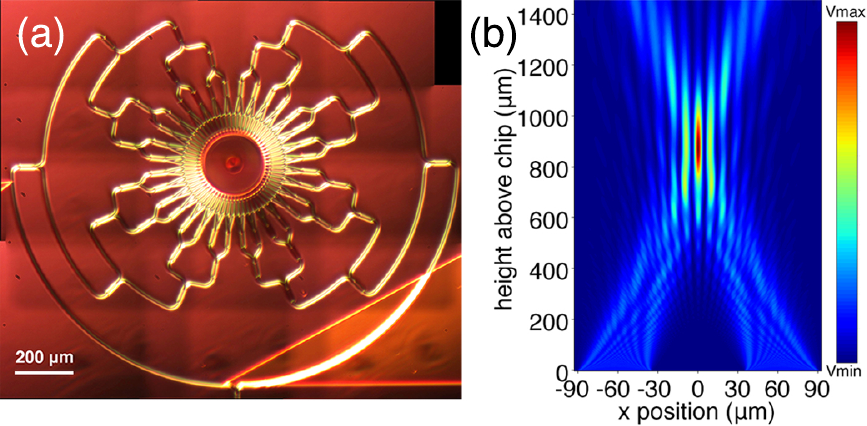}
\caption{(a) microscope image (stitched) of the fabricated device. (b) simulation results using two plane waves of sizes and field profile equivalent to the chip based measurement to show how an axicon may behave.}
\label{fig:device-image}
\end{figure}

% Materials: fabrication
The device is a SOI chip that consists of a 220$\pm$20 nm Silicon layer atop a 2 $\mu$m $SiO_2$ BOX layer, which is on a Silicon substrate. The chip also consists of a 1 $\mu$m protective $SiO_2$ layer. All waveguides are rib waveguides etched 120$\pm$10 nm, and the gratings are etched 70$\pm$10 nm from the surface. The reported waveguide losses from the fabricators was $\approx$4 dB/cm. The chips were fabricated using DUV lithography in the 10th multi-wafer project call by the CORNERSTONE Project~\cite{cornerstone10}. The feature sizes were limited by the fabrication parameters, with a minimum feature size of 200 nm and gap of 250 nm in the relevant gratings layer. This coarse limit, compared to $\approx$25 nm resolutions available in the present day DUV lithography techniques and electron-beam lithography, prevents us from designing higher performance grating structures with the usual apodization in the radial direction. However, we were still able to design an axicon while staying within the fabrication limits.

% Materials: chip image
The axicon, shown in Fig.~\ref{fig:device-image}a, is a 1.52 x 1.38 mm$^2$ device that consists of seven levels of 1x2 MMIs to combine or split light (depending on the coupling direction) which leads to 128 final ports around the central axicon structure. The central structure of gratings is $\approx$182 $\mu$m in diameter, and has a period of 600 nm and a duty cycle of 0.5. 

% Methods: axicon theory
In a diffraction grating-based axicon, we design the device such that light would diffract from all sides towards the center, and a central lobe would form due to constructive interference at the center. In order to show how such an axicon would behave, we performed a 2D-FDTD simulation with two plane waves at a wavelength of 1310 nm. We set the angle of the two sources, as well as the width of the individual source, based on the measurement results (discussed in detail in Sec.~\ref{sec:results}\ref{sec:results-apodization} and \ref{sec:results}\ref{sec:results-power}). Figure~\ref{fig:device-image}b shows the electric field intensity of the simulation results. Light from the two sources interfere constructively to form a central lobe that spans from $\approx$750 $\mu$m to 1000 $\mu$m above the chip, with a lobe diameter of $\approx$7 $\mu$m.

%With the gratings of constant period, light at a particular wavelength diffracts at a fixed angle. With a broadband source, the diffracted light would have a range of angles. We performed 2D-FDTD simulations \textit{using ANSYS/Lumerical}) show that the incoming light at 1275 nm would diffract at 45$^{\circ}$ from the vertical, and the angle would decrease to 35$^{\circ}$ at a wavelength of 1350 nm, providing a multi-spectral lobe length of \textbf{enter from simulation}. While these numbers are do give us an idea of the qualitative performance, the real values are expected to be fairly different due to fabrication variances. % For such a source, we would expect the central lobe to span \textbf{EXPECTED LOBE LENGTH} against a device diameter of 182 $\mu$m. 

%Laser and measurement system

\begin{figure}[htbp]
\centering
\includegraphics[width=0.75\linewidth]{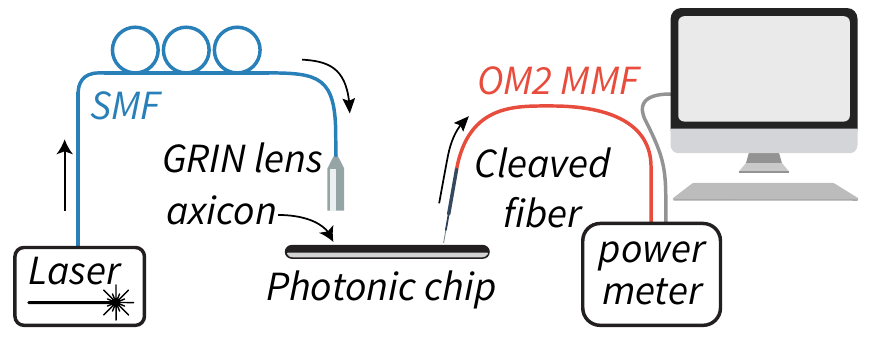}
\caption{Schematic of the measurement setup for power characterization. The arrows show the direction of light propagation.}
\label{fig:powermeter-setup}
\end{figure}

% reverse measurement, polarization
Figure~\ref{fig:powermeter-setup} shows the experimental setup used to characterize the performance of the device. For the source, we used a \textit{Santec HSL-20} swept source 1300 nm laser with a bandwidth of 100 nm. Instead of coupling light in from a grating coupler and measuring the output light at the axicon, as is typical in these kinds of devices, we did the reverse. We coupled light in from the axicon structure, and measured out from the grating coupler. Light was coupled in with a GRIN lens (NA 0.46, \textit{Thorlabs GRIN2913}) mounted on a ferruled fiber and held vertically over the chip. The output light from the grating coupler was coupled to a cleaved OM2 multimode fiber (MMF) (core diameter 50 $\mu$m, NA 0.20). The power was recorded with a power meter \textit{(Thorlabs PM100D)} mounted with an \textit{InGaAs} detector \textit{(Thorlabs S154C)}. 

We first optimized the polarization of the light with the manual paddle polarizers (27 mm diameter paddles, with 4-3-4 loops) and the power meter readings. We then moved the GRIN lens point-by-point using a motorized $xyz$ translation stage \textit{(Thorlabs NanoMax313D)} to acquire planar data from the chip. Using a MMF to record the output light benefits from the larger NA of the MMF, which allows us to accept light from a larger acceptance cone accounting for more angles out of the grating coupler which ultimately results in a higher spectral bandwidth. We note that we did not adjust the phase of different paths in the device.

% Results
\section{Results}
\label{sec:results}

\subsection{Effect of azimuthal apodization}
\label{sec:results-apodization}

In regular gratings, the intensity of light diffracted decreases exponentially with penetration depth. For a large gratings based device such as the axicon presented here, the majority of the light would diffract in the first few grating periods. An axicon by design requires light to penetrate much deeper into the device, with a uniform intensity, as depicted in Fig.~\ref{fig:device-image}b. In order to increase that penetration depth so that a longer and narrower depth of focus can be achieved, we broke up the circular gratings into small arcs, i.e. we azimuthally apodized the gratings, as described in a previous work~\cite{Maharjan2020}. The width of the central lobe depends on the penetration depth by their Fourier transform by virtue of the Fraunhauffer diffraction relationship~\cite{BornWolfChapVIII1999}. The arc-length of the gratings are shortest at the circumference and increases exponentially with penetration depth. We obtained the extinction coefficient for the apodization from 2D-FDTD simulations. The arcs continued until the gap between two arcs at a given radius reached the feature-size limit for fabrication, which was at a radius of 28 $\mu$m in our device. This apodization is designed to change the diffraction intensity as a function of penetration depth from exponential decay to nearly a constant. 

% apodization

\begin{figure}[htbp]
\centering
\includegraphics[width=0.95\linewidth]{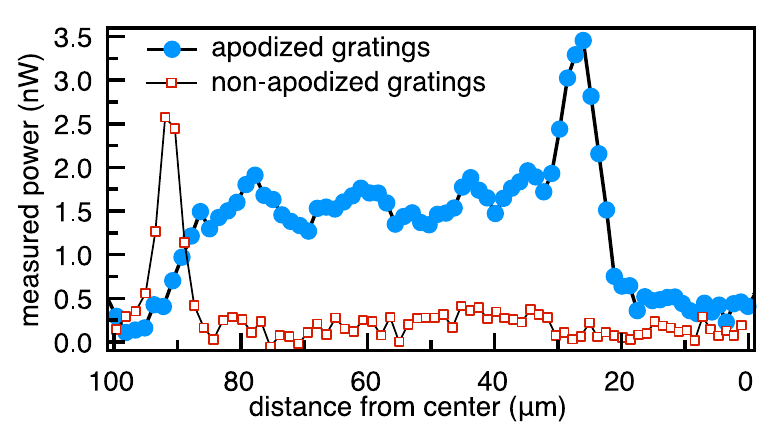}
\caption{Power measured at central location of the device for apodized and non-apodized gratings. Azimuthally apodized gratings allow for deeper penetration of light, to $\approx$55 $\mu$m, while the non-apodized gratings fall off exponentially within $\approx$5 $\mu$m due to rescattering.}
\label{fig:apodization}
\end{figure}

Figure ~\ref{fig:apodization} shows the measured power for the apodized (blue circles) and the non-apodized (red squares) gratings for the axicon as a function of radius of the axicon (distance from the center). For non-apodized gratings, we see that almost all of the light is diffracted away within the first few gratings near the outer radius of the device, and the penetration depth is $\approx$5 $\mu$m. In contrast, for the apodized-gratings based device, we see that light makes it through most of the device. The diffraction intensity is nearly constant for the first $\approx$55 $\mu$m, at which point the full circular gratings start, and we observe the intensity increase and then decrease to zero. With a high-resolution fabrication technique, we would likely be able to flatten that spike in intensity and increase the penetration depth past the current $\approx$55 $\mu$m.

\subsection{Power characterization}
\label{sec:results-power}
\begin{figure}[htbp]
\centering
\includegraphics[width=0.9\linewidth]{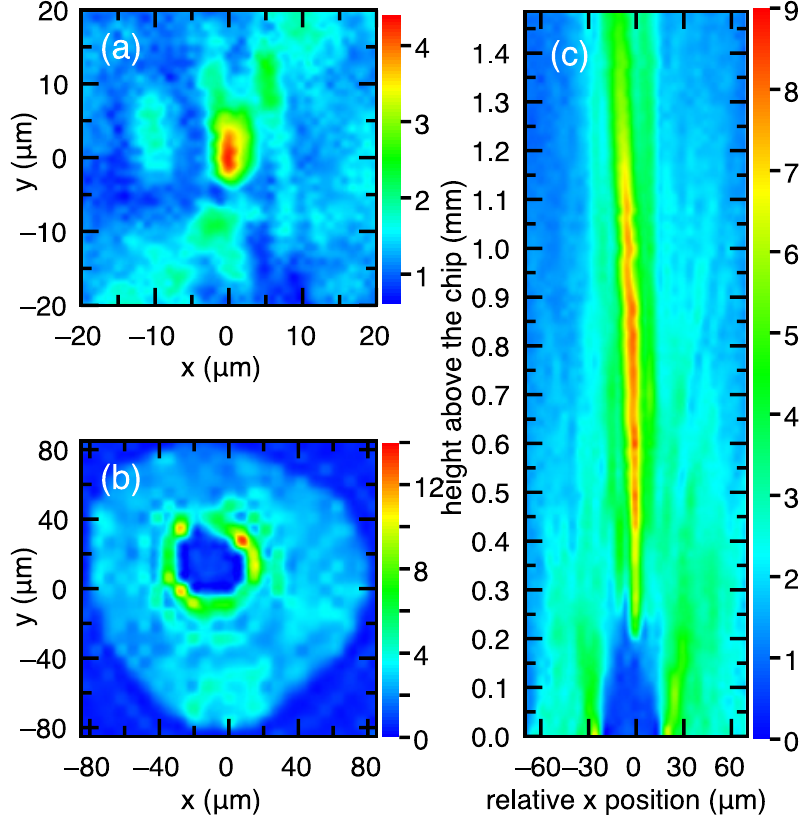}
\caption{Power measurement of the axicon. Sub-figure (a) shows the $xy$-plane at a height of 400 $\mu$m above the chip, (b) shows the $xy$-plane at the chip surface, and (c) shows the $xz$-plane at the central lobe $y$ position. All color bars are in units of nW.}
\label{fig:axicon-power}
\end{figure}

With the technique described in Sec.~\ref{sec:methods}, we measured $xy$- and $xz$-planes for the device with the power meter. Figure~\ref{fig:axicon-power} shows the two planes of a typical measurement. In Fig.~\ref{fig:axicon-power}b, we present the power measured at the $xy$-plane at the surface of the chip. We started the measurement from the most positive $x$ position $(x_{max})$ at the most negative $y$ position $(y_{min})$ and moved $x$ in a fixed step size until the most negative position $x_{min}$, and retracted back to $x_{max}$. We only recorded power data in one direction to minimize any errors due to mechanical backlash in the translation stages. We then moved $y$ at the fixed step size and repeated until we reach $(y_{max})$. For Fig.~\ref{fig:axicon-power}b, we used a step size of 10 $\mu$m. We note that the shape is not quite circular due to mechanical effects in the translation stage. We used a dial test indicator to measure the difference between true linear movement versus the rotation of the micrometer dials in the translation stage, and found that the true movement was larger than the micrometer dial step size at near $x_{max}$ and $y_{min}$ positions, and the true movement was smaller than the step size near $x_{min}$ position and $y_{max}$. Figure~\ref{fig:axicon-power}a shows a similar $xy$-plane measurement but performed at a height of 400 $\mu$m above the chip surface. This plane was acquired at a much higher step size of 1 $\mu$m. We fitted a gaussian function to this data to find that lobe width at FWHM was 7.6 $\mu$m.

Based on results of planar data at the surface, we identify the central $y$ location, which allows us to acquire a $xz$-plane, one example of which is shown in Fig.~\ref{fig:axicon-power}c. The step size in $x$ was 5 $\mu$m and $z$ was 15 $\mu$m. We note that the the $z$ zero position has an uncertainty of 25 $\mu$m. From Fig.~\ref{fig:axicon-power}c, we can see that the central lobe extends from $\approx$350 $\mu$m to $\approx$1200 $\mu$m.

\subsection{Spectral characterization}
\label{sec:results-spectral}

While Fig.~\ref{fig:axicon-power} shows the absolute power measurements, we also analyzed the spectral performance of the axicon, within the range that our source laser could provide, and at a fixed grating angle at the output. The power meter measurements do not need very high speed data acquisition rate, but faster systems were needed in order to resolve the wavelength. We used \textit{AlazarTech ATS9350} data acquisition (DAQ) card and a balanced detector \textit{(Thorlabs PDB430C)} to acquire data at 500MHz for the spectral measurements. However, the spectral power density out from the chip was in the order of the noise floor of the detector, so we implemented a balanced homodyne detection method to acquire the signal. Since this setup requires us to mix two signals, all the fibers used in this measurement are single mode fibers.

% homodyne detection block diagram/schematic
\begin{figure}[htbp]
\centering
\includegraphics[width=0.9\linewidth]{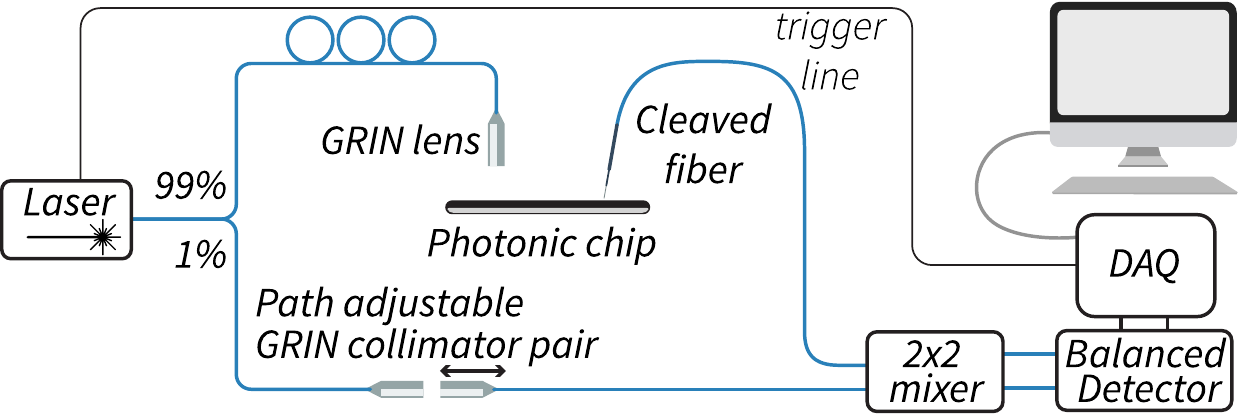}
\caption{Schematic of the setup for balanced homodyne detection to perform spectral analysis.}
\label{fig:balanced-homodyne}
\end{figure}

Figure~\ref{fig:balanced-homodyne} shows the setup for our implementation of balanced homodyne detection based measurements. The source laser is first split using a commercial 1:99 splitter. The 99\% line goes through the polarization rotator, and out of the assembled GRIN coupler to couple light onto the chip. Light out of the chip is coupled onto a cleaved fiber held at a fixed angle. The cleaved fiber is mounted on a manual $xyz$-stage and the GRIN lens is mounted on a motorized $xyz$-stage. The 1\% line is the local oscillator, which has a pair of GRIN collimators \textit{(Thorlabs 50-1310A-APC)} to adjust the path length. One of the GRIN lens is mounted on a motorized linear translation stage, which allows us to increase or decrease the path length as needed. This is specially important when we acquire data of a plane perpendicular to the chip. As the input GRIN moves away from the chip, the path length increases. Having a synced motor to change the path length by the corresponding distance on the local oscillator line would maintain that equality in path length. The light out of the chip is then mixed with the light from the local oscillator line in a 2x2 mixer, which feeds into the balanced detector. The voltage data from the balanced detector is then acquired using the DAQ card.

% homodyne theory
If we assume two signals $A_s\sin{(\theta + \phi)}$ and $B_{LO}\sin{\theta}$ entering the 2x2 mixer, then the output of the mixer is $(A_s/\sqrt{2})\cos{(\theta + \phi)} - (B_{LO}/\sqrt{2})\sin{\theta}$ and $(-A_s/\sqrt{2})\sin{(\theta + \phi)} + (B_{LO}/\sqrt{2})\cos{\theta}$. The photo-diodes in the balanced detector measure the intensities, and the balanced detection is the difference of the two intensities. Thus, we obtain the difference of the intensities
\begin{equation}
    \frac{A_s^2}{2}\cos{2(\theta + \phi)} - \frac{B_{LO}^2}{2}\cos{2\theta} + A_s B_{LO}\sin{\phi}.
\label{eq:difference-intensity}
\end{equation}
Let us assume $A_s\sin{(\theta + \phi)}$ and $B_{LO}\sin{\theta}$ are the chip signal and the local oscillator, respectively. Correspondingly, $A_s^2/2$ and $B_{LO}^2/2$ would represent their intensities and since $A_s\ll B_{LO}$, $A_s^2$ would be negligible compared to $B_{LO}^2$. When the path lengths of both signals are equal, $A_s B_{LO}$ would represent the constructive interference component. So, if we subtract the $B_{LO}^2$ component from (\ref{eq:difference-intensity}), we can isolate the interference component and extract the chip signal.

% homodyne detection algorithm
\begin{figure}[htbp]
\centering
\includegraphics[width=0.9\linewidth]{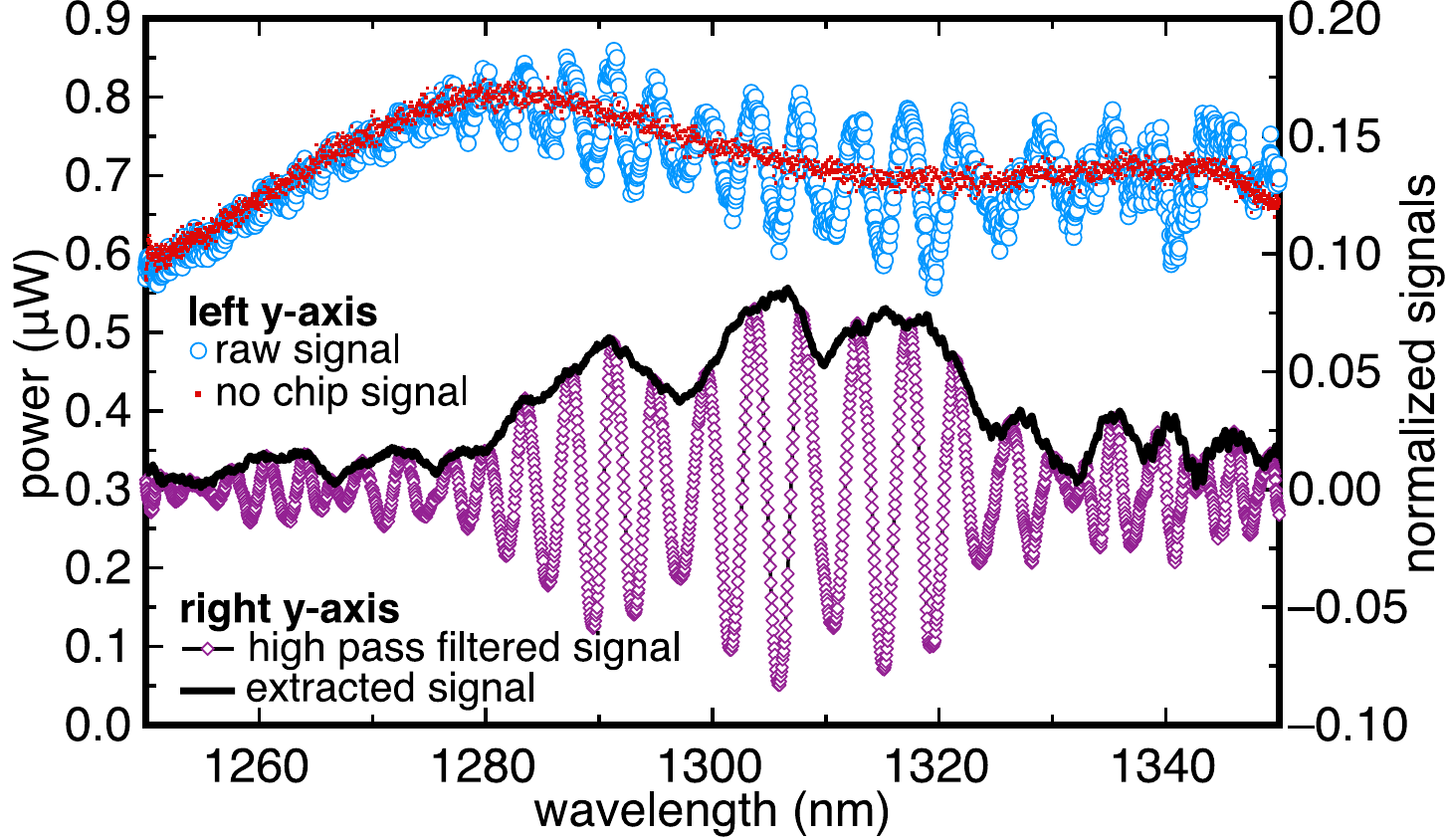}
\caption{Representative example of a homodyne detection algorithm implementation.}
\label{fig:homodyne-algorithm}
\end{figure}

Figure~\ref{fig:homodyne-algorithm} shows an example implementation of this algorithm to extract the chip signal from the homodyne setup. On the left axis, we have the raw signals. Prior to beginning the experiment, we record a signal (average of 50 measurements) without the chip connected. This is the \textit{no chip signal} in Fig.~\ref{fig:homodyne-algorithm}. When the chip is connected, the local oscillator and the chip signal is mixed in the 2x2 mixer. If the two path lengths of the chip signal and the local oscillator signal are nearly equal, then we can observe an interference pattern in the balanced detection signal (difference of the two signals). This is the \textit{raw signal} in Fig.~\ref{fig:homodyne-algorithm} (average of 25 measurements). The envelope of the interference pattern is the component representing the signal from the chip. 

To extract the desired chip signal, we first smooth the \textit{raw signal} and \textit{no chip signal} over 60 ns, and normalize it with their individual maxima. We take the difference of the two signals to extract only the interference component, which we pass through a high pass filter, filtering out any signal lower than 5MHz. This leaves us with \textit{high pass filtered signal} in Fig.~\ref{fig:homodyne-algorithm}. We then implement the Hilbert function to extract the envelope of the oscillation, which is the component from the chip, shown as \textit{extracted signal} in Fig.~\ref{fig:homodyne-algorithm}. Since we use a swept source laser, we can extract different sections at the required wavelengths to study the spectral dependence.

\begin{figure}[htbp]
\centering
\includegraphics[width=0.9\linewidth]{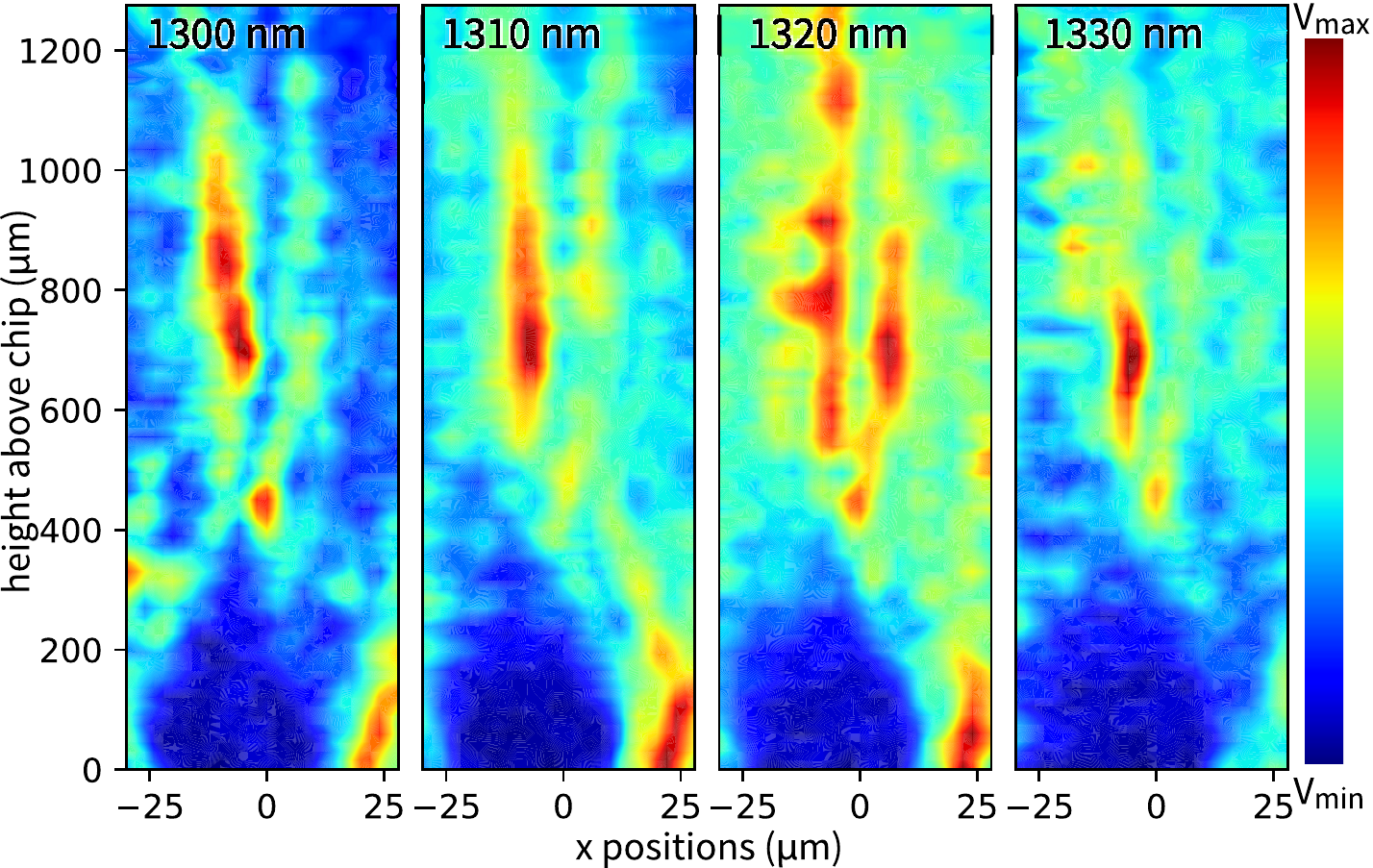}
\caption{Spectral characterization of the axicon. The four panels present the power at the wavelengths shown at the top. The bandwidth used for each plot is 10 nm. The values are normalized to each maxima.}
\label{fig:axicon-spectral}
\end{figure}

Figure~\ref{fig:axicon-spectral} shows the spectral behavior of the axicon. The data, acquired at a resolution of 2 $\mu$m in $x$ and 15 $\mu$m in $z$, is limited to a small region of 60 $\mu$m near the central lobe in order to keep the experiment short enough that the fibers do not get misaligned. Since we performed homodyne detection to acquire the signal in this case, we also normalized the measured values. We find that the focal lobe of the axicon is relatively stable, and does not vary significantly with respect to the distance from the chip at different wavelengths discussed in Fig.~\ref{fig:axicon-spectral}.

\section{Conclusions}
\label{sec:conclusions}

We demonstrate a chip-based axicon developed with a low-resolution fabrication technique that limits the design to 200 nm feature sizes, and gaps of 250 nm. We employed azimuthal apodization by breaking up the circular gratings to increase the penetration depth of the light, from $\approx$5 $\mu$m to $\approx$55 $\mu$m, and also changed the diffraction intensity profile from exponential to nearly constant. We characterized this axicon by coupling light in through the axicon itself with a GRIN lens and measuring the combined light via a grating coupler with a multimode fiber using a power meter. The multimode fiber has a larger acceptance angle and therefore, a larger spectral bandwidth can be measured. Using this setup, we measured the axicon to have a central lobe that is 7.6 $\mu$m in diameter, and of length $\approx$850 $\mu$m. Further, we implemented homodyne detection in order to perform a spectral analysis and determined that the focus of this axicon does not significantly change its axial position at different wavelengths tested, from 1300 nm to 1330 nm. A device like this, with the longer depth of focus and wavelength independence, could be beneficial in various applications such as optical coherence tomography.

\section{Funding}
This work is supported by the GCRF program of the EPSRC under Grant No. EP/R014418/1, and partially supported by TWAS Grant No. 18-013RG/PHYS.

\section{Acknowledgments}
We thank ANSYS/Lumerical for the simulation software, Luceda Photonics for the IPKISS chip design software, and Thorlabs for various optomechanical hardware. We also thank the CORNERSTONE Project at the University of Southampton for fabricating the chips.

\noindent\textbf{Disclosures.} The authors declare no conflicts of interest.

% Bibliography

\end{document}